\documentclass[a4paper,11pt]{article}
\pdfoutput=1 

\usepackage{jcappub} 

\usepackage[T1]{fontenc} 
\usepackage{natbib}
\setcitestyle{square,numbers}
\usepackage{siunitx}
\usepackage{lineno}
\usepackage{xparse}
\ExplSyntaxOn
\NewDocumentCommand{\longdash}{ O{2} }
 {
  --\prg_replicate:nn { #1 - 1 } { \negthinspace -- }
 }
\ExplSyntaxOff

\notoc
\title{Deep learning for improved keV-scale recoil identification in high resolution gas time projection chambers}

\author[a,1]{J.~Schueler,\note{Corresponding author.}}
\author[a]{M.~Ghrear,}
\author[a]{S.~E.~Vahsen,}
\author[b]{P.~Sadowski,}
\author[c]{C.~Deaconu}

\affiliation[a]{Department of Physics and Astronomy, University of Hawaii, Honolulu, Hawaii 96822, USA}
\affiliation[b]{Information and Computer Sciences Department, University of Hawaii, Honolulu, Hawaii 96822, USA}
\affiliation[c]{Department of Physics, Enrico Fermi Inst., Kavli Inst. for Cosmological Physics, University of Chicago, Chicago, IL 60637, USA}

\emailAdd{jschuel@hawaii.edu}
\emailAdd{majd@hawaii.edu}
\emailAdd{sevahsen@hawaii.edu}

\abstract{Recoil-imaging gaseous time projection chambers (TPCs) with directional sensitivity are attractive for dark matter (DM) searches. Detectors capable of reconstructing 3D nuclear recoil directions would be uniquely sensitive to the predicted dipole angular distribution of DM recoils. Observation of this directional distribution would unambiguously establish the galactic origin of a claimed DM signal. Recoil directionality also would provide powerful discrimination against background recoils from solar neutrino scattering. These advantages of directionality can only be exploited however, if electron recoil backgrounds from gamma rays can be sufficiently suppressed. We introduce a deep learning-based recoil event classifier that uses a 3D convolutional neural network (3DCNN) to identify event species based on their recoil images. We compare electron background rejection performance of discriminants determined by the 3DCNN both to the traditional discriminant of track length, as well as discriminants obtained from state-of-the-art shallow learning methods. We train the 3DCNN classifier using recoil charge distributions with ionization energies ranging from 0.5-\SI{10.5}{keV_{ee}}, for \SI{25}{cm} of drift in an 80:10:10 mixture of $\rm He$:$\rm CF_4$:$\rm CHF_3$. The charges are initially segmented into $(100\times 100\times 100)$\SI{}{\um^3} bins when determining track length and the shallow learning discriminants, but are rebinned with a reduced segmentation of about $(850\times 850\times 850)$\SI{}{\um^3} for the 3DCNN. Despite the courser binning, compared to using track length, we find that classifying events with the 3DCNN reduces electron backgrounds by a factor of up to 1,000 and effectively reduces the energy threshold of our simulated TPC by $30\%$ for fluorine recoils and $50\%$ for helium recoils. We also find that the 3DCNN reduces electron backgrounds by up to a factor of 20 compared to the shallow machine learning approaches, corresponding to a \SI{2}{keV_{ee}} reduction in the energy threshold.}

\begin{document}
\maketitle
\flushbottom

\section{Introduction}
\label{sec:intro}

Directly detecting the constituents of dark matter (DM) remains one of the key goals of contemporary physics. Directional detection would provide unique, robust, and unambiguous confirmation of the galactic origin of a signal in the form of a dipole distribution in galactic coordinates~\cite{spergel, Mayet:2016zxu}.

As more regions of DM parameter space are being ruled out, and cross-section limits approach the solar neutrino fog, directional detection is also increasingly of interest for maximizing DM sensitivity within the fog, and for studying the solar neutrinos themselves. Detectors capable of reconstructing the full 3D vector direction, energy, and time of individual recoil events are preferable in this context. Such ``recoil-imaging'' detectors are widely applicable and hence garnering increasing interest~\cite{Vahsen:2021gnb, OHare:2022jnx}.

Gaseous time projection chambers (TPCs) are the most mature recoil imaging technology with event time measurement. Unlike liquid noble gas detector, which benefit from strong self-shielding, one key issue in gas detectors is the rejection of electron recoil backgrounds from gamma rays. TPCs with high-definition charge readout (HD TPCs) are, however, capable of reconstructing the topology of both nuclear recoils and electron recoils in great detail, which we expect should maximize particle ID capabilities and again help to reject such backgrounds. A key question then, is whether the electron rejection in gas TPCs is sufficient to achieve a background free DM search for a given experiment and exposure.

Studies of the proposed \SI{1000}{m^3} gas TPC experiment CYGNUS \cite{cygnus}, running for six years, suggest the electron background will be flat in energy, and electron backgrounds must be rejected offline by factors exceeding $6\times 10^4$ per \SI{}{keV_{ee}}. Due to the diffusion in gas-based detectors, electron identification performance falls exponentially with recoil energy. As a result, the effective analysis energy threshold above which a large gas TPC will remain background free is determined by the particular energy at which the electron rejection meets the performance requirement. A relatively small improvement in electron rejection, and consequent reduction of the analysis energy threshold can have large benefits on DM reach, because the expected DM recoil spectrum is generally steeply falling with energy.

Because recoil-imaging HD TPCs offer extremely rich three-dimensional charge density measurements, one obvious question is how to best exploit this information to maximize recoil identification. This is a problem ideally suited for machine learning techniques.

There have been many efforts to improve electron background rejection for keV-scale recoil events, which often use multivariate combinations of discriminant observables for event classification. The MIMAC group, for instance, used boosted decision trees (BDTs) \cite{bdt} to improve electron rejection factors by a factor of 20 over more traditional methods \cite{billard}. More recently, we introduced a set of nine observables that are based on the shape of the recoil charge cloud measured in a 3D recoil-imaging TPC \cite{ghrear}. A joint combination of these nine event-shape variables using a hard-cuts-based approach led to up to a two order of magnitude improvement in electron rejection over using the length along a track's principal axis, which is a common observable for event classification.

Given the electron background rejection improvement observed both by multivariate combinations of discriminant variables and the usage of event-shape variables, we attempt to combine the best aspects of both of these approaches by introducing a deep learning-based classifier for event identification. In particular, we introduce a 3D convolutional neural network (3DCNN, Section \ref{sec:cnn}) that is directly fed the 3D ionization density distribution of events, binned into a $32\times 32\times 32$ voxel grid, and outputs class probabilities of the recoil species of the event. We expect this end-to-end approach to enable us to capture more information than using predefined observables \cite{baldi2014searching, baldi2016jet, sadowski2017efficient, sadowski2018deep}, thus leading to better background rejection performance. In Section \ref{sec:erej}, we separately combine the nine event-shape observables with two shallow learning techniques, (1) a BDT and (2) a fully connected neural network (FCNN), both of which also output recoil-species class probabilities that can be used as multivariate classification discriminants. Comparing the electron background rejection performance between these shallow learning-produced discriminants and the hard-cuts-based combined observable from Ref. \cite{ghrear} allows us to assess the relative effectiveness of different techniques of combining the nine event-shape observables. Comparing the electron rejection performance of the 3DCNN with the best-performing combination of these observables provides insight on if any crucial event-shape information is missing from the set of nine observables.

\section{Overview of simulation}

Before introducing our new classifiers, we first describe the simulated detector and recoil characteristics used in our study. We build off of previous work from our group and generate a large simulation sample using identical parameters to those in Ref. \cite{ghrear}. Doing this allows us to directly compare the electron rejection performance of our new classifiers with already established electron rejection improvements.

We use \texttt{SRIM} \cite{srim} and \texttt{retrim} \cite{retrim} to simulate recoiling He and F nuclei and \texttt{DEGRAD} \cite{degrad} to simulate electron recoils in an 80:10:10 mixture of $\text{He}$:$\text{CF}_4$:$\text{CHF}_3$ at a total pressure of \SI{60}{Torr} and temperature of $25^{\circ}$C. The \texttt{SRIM} computation requires a compound correction for every nucleus in the gas mixture (He, H, C, and F). The calculation of these compound corrections and their associated values are detailed in Ref. \cite{ghrear}. The \texttt{retrim} step requires the average energy per ion pair, $W$, and Fano factor, $\mathcal{F}$, which were calculated using \texttt{Garfield++/Heed} \cite{garfield} as $W = \SI{35}{eV}$ and $\mathcal{F} = 0.19$, respectively. \texttt{DEGRAD} can simulate recoils isotropically; however, for \texttt{SRIM} and \texttt{retrim} we isotropize the recoil simulations after generating the initial ionization distributions. We assume a drift field of \SI{40.6}{V/cm} parallel to the drift direction and use \texttt{Magboltz} \cite{magboltz} to determine the corresponding transverse and longitudinal diffusion coefficients as $(\sigma_T,\sigma_L) = (398,425)$\SI{}{\um/\sqrt{cm}}. The diffusion is applied to our simulations assuming a uniform drift length of \SI{25}{cm}.

After applying diffusion, we assume each individual electron is detected. We do not simulated any charge amplification or digitization, but instead simulate a highly efficient, pixelated readout by binning each primary track into $(\SI{100}{\um}\times \SI{100}{\um}\times \SI{100}{\um})$ voxels. We note that our choice of a \SI{40.6}{V/cm} drift field corresponds to $\SI{100}{\um}$ per clock cycle on a $\SI{40}{MHz}$ clock which is equivalent to the readout refresh rate used in existing TPCs with pixel readout \cite{jaegle}.

\section{Convolutional neural network classifier}
\label{sec:cnn}
\begin{figure}[htbp]
\centering 
\includegraphics[width=\textwidth]{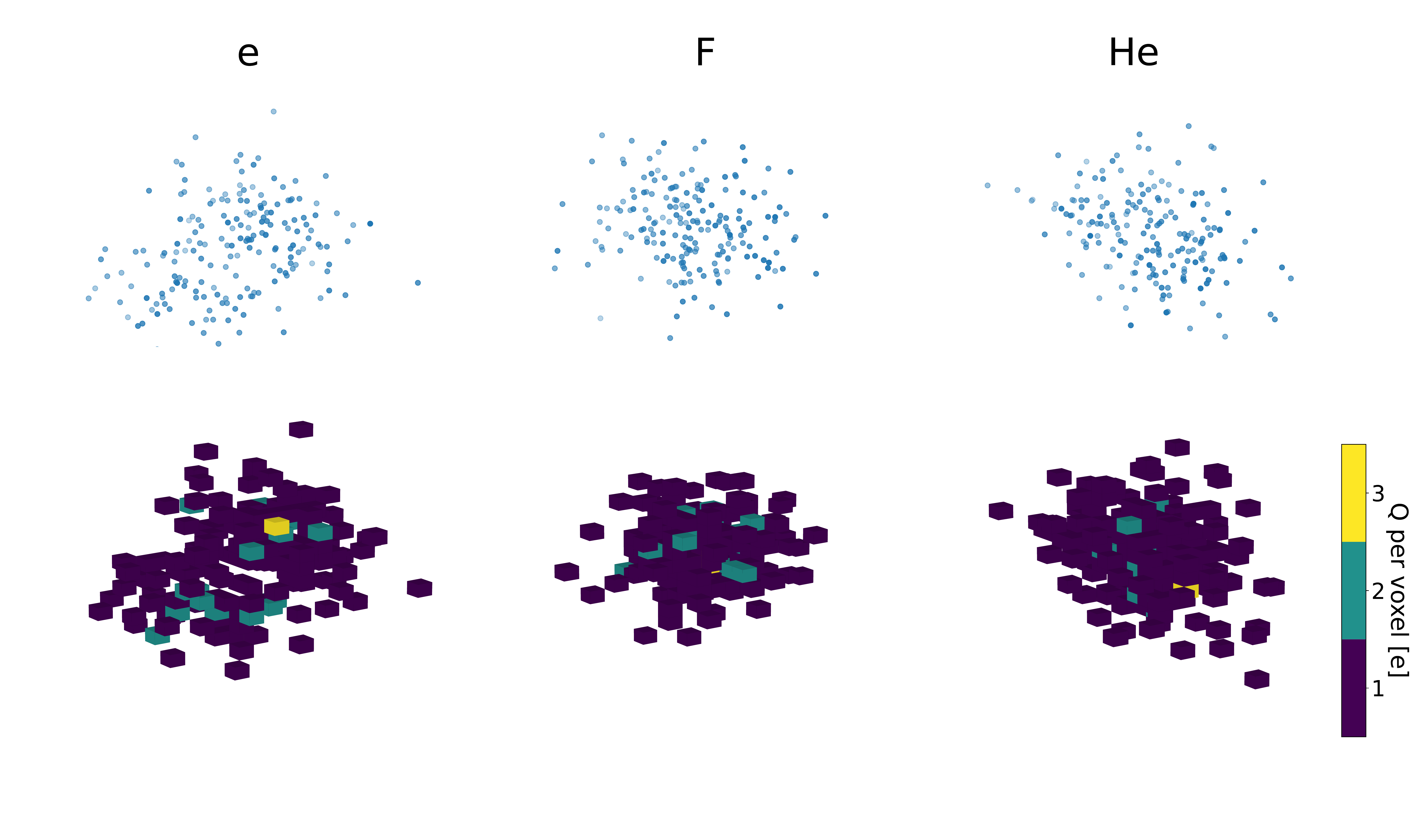}
\caption{\label{fig:0} 3D charge distributions of \SI{6}{keV_{ee}} electron, fluorine, and helium recoils after diffusion. The top row shows the original $(\SI{100}{\um}\times \SI{100}{\um}\times \SI{100}{\um})$ binning of these events, which is fine enough that there is only a single charge in each filled bin. The bottom row shows the events re-binned into the $(32\times 32 \times 32)$ voxel grid that are input into the 3DCNN.}
\end{figure}

Recoil-imaging TPCs with high readout segmentation reconstruct detailed 3D images of ionization charge distributions of recoil tracks. These tracks have many identifiable characteristics that can be used for particle identification. For sufficiently high energy recoil tracks, the length of the track along its principal axis is often enough to reliably distinguish between electron recoils and nuclear recoils, as electron recoils tend to create longer tracks following a more meandering path than nuclear recoils of equivalent energy. At lower energies, however, diffusion during drift has a proportionately larger effect on the overall event topology, leading to more spherical ionization distributions. Figure \ref{fig:0} shows $\sim$\SI{6}{keV_{ee}} examples of each of the three recoil species investigated here. Comparing the background electron recoil (left) with the signal fluorine and helium recoils (middle and right, respectively), we see that both of the nuclear recoil species have a dense cluster of charge near the center of the event, whereas the electron recoil appears to be a more diffuse charge cloud with larger gaps. It's difficult to unambiguously identify the principal axis direction of low energy tracks, leading to poor angular resolution. Furthermore, a few stray charges can significantly bias the length along the principal axis, which may lead to event misclassification with this traditional approach. Since there are so many features present in these tracks, using a deep learning classifier where the classifier learns the best patterns for event selection on its own, independent of an identified principal axis, is attractive. To this end, we construct and train a 3DCNN classifier that is directly fed 3D voxel grids of the ionization distribution of individual recoil events with the recoil species assigned as the class label of the event.

\subsection{Building blocks of feature extraction in 3DCNNs}
\begin{figure}[htbp]
\centering 
\includegraphics[width=0.6\textwidth]{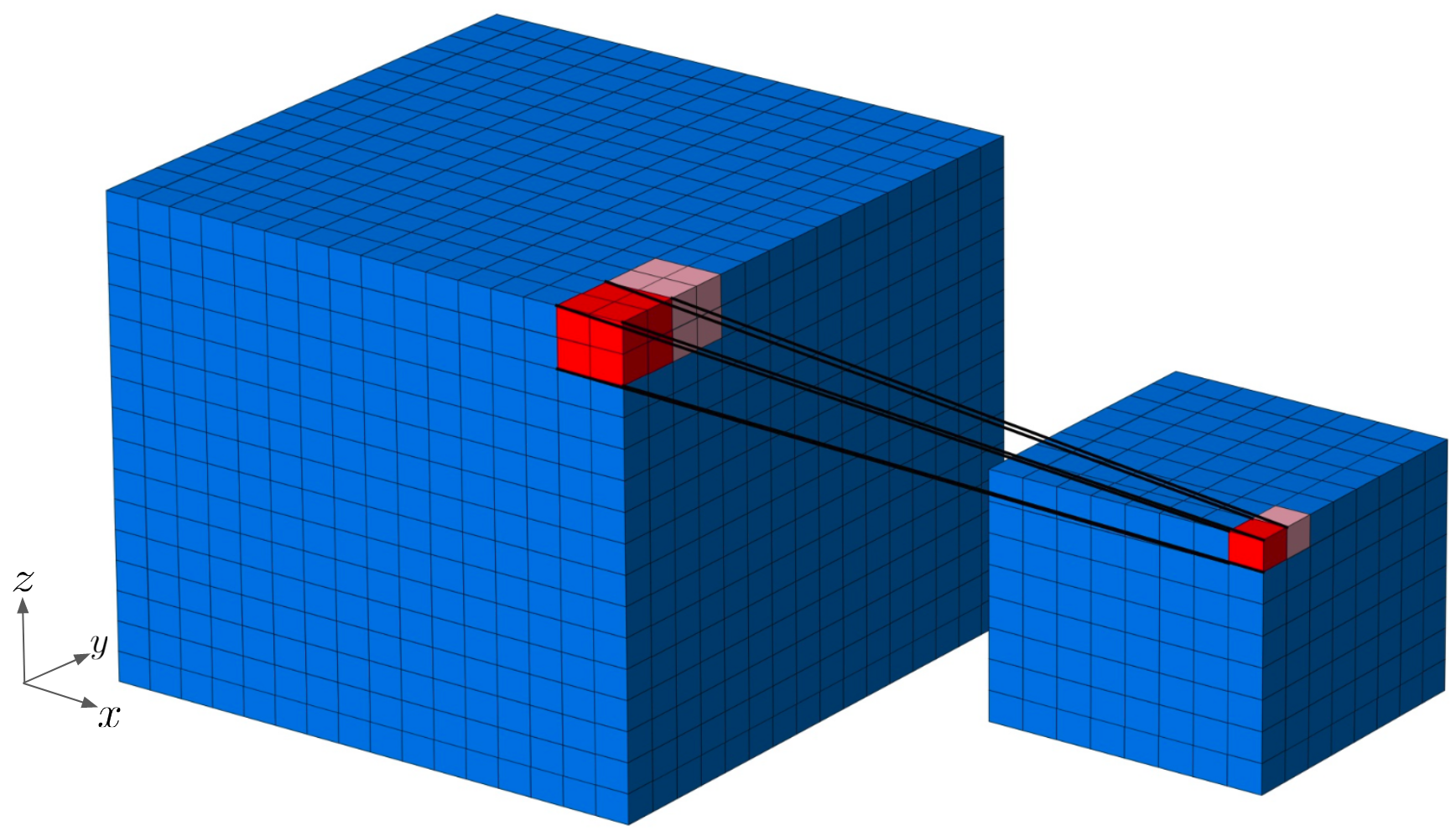}
\caption{\label{fig:2} Schematic representation of the dimensionality reduction of a 3D convolution operation with a convolutional stride of $S = (2,2,2)$. Here $N\in\mathbb{R}^{16\times 16\times 16}$, and $K\in\mathbb{R}^{2\times 2\times 2}$. The black lines connecting the two red shaded regions illustrate the action of $K$ on $N[14,0,14]$, a $2\times 2\times 2$ sub-block of $N$, leading to the $1\times 1\times 1$ element $C_{14,0,14}$ of $C\in \mathbb{R}^{8\times 8\times 8}$. Each index $\alpha$, $\beta$, and $\gamma$ of $C$ runs along $\{0,2,4,6,8,10,12,14\}$ due to the convolutional stride of $S = (2,2,2)$. The pink regions show the next element of the convolution operation after striding 2 units along the $y$ direction on $N$.}
\end{figure}
3DCNNs are not yet widely used for natural image classification due to the large amounts of GPU memory required to store and perform operations on large 3D images with high voxel density. The lower energy recoil events that are of interest in directional DM searches, however, have their entire event topologies contained within a relatively small region, enabling the use of 3DCNNs with current-generation hardware, even in detectors with high spatial resolution. Indeed, since convolutional neural networks are translationally invariant, given a set of 3D recoil tracks, we can shift the coordinates of each recoil track to a fixed origin without any change in performance of a 3DCNN up to uncertainties in bin assignment. Translating events to a common origin allows for the fiducial binning of track charges into a much smaller 3D grid of voxels than binning the entire fiducial volume of a detector. Here we briefly introduce 3D convolutions and 3D pooling; two core building blocks of our 3DCNN.
\subsubsection{3D Convolutions}
\label{subsubsec:3dconv}

In 3DCNNs, the 3D convolution operation is a 3D cross correlation of an input grid of data, $N$, with a kernel (also called a filter), $K$, that slides along the input grid. To be more specific, let $N\in\mathbb{R}^{n_x\times n_y\times n_z}$ (often called an input feature map), and let $K\in\mathbb{R}^{k_x\times k_y\times k_z}$, where $k_j\leq n_j;\quad j=x,y,z$. We denote the shapes of $N$ and $K$ as $(n_x\times n_y\times n_z)$ and $(k_x\times k_y\times k_z)$, respectively. The action of $K$ on $N[\alpha,\beta,\gamma]$, a $k_x\times k_y\times k_z$ sub-block of $N$, is an element, $C_{\alpha\beta\gamma},$ of the output ``convolved'' feature map, $C$, and is a mapping from $\mathbb{R}^{k_x\times k_y\times k_z}\rightarrow\mathbb{R}$ defined by

\begin{align}
\label{eq:1}
    (K\star N[\alpha,\beta,\gamma])_{\alpha\beta\gamma} = \sum_{l=0}^{k_x-1}\sum_{m=0}^{k_y-1}\sum_{n=0}^{k_z-1}N_{\alpha+l,\beta+m,\gamma+n}K_{lmn} \equiv C_{\alpha\beta\gamma},
\end{align}
where $\alpha+k_x \leq n_x$, $\beta+k_y\leq n_y$, $\gamma+k_z\leq n_z$, and $(K\star N[\alpha,\beta,\gamma])_{\alpha\beta\gamma}$ is a single element of the 3D cross correlation operation. 
 
$C$ is formed by sliding $K$ in integer steps of $S_x$, $S_y$ and $S_z$ along the $x$, $y$, and $z$ extents of $N$, respectively. In particular, $C$ is an ordered grid composed of all \newline $\lfloor\frac{n_x-k_x}{S_x}+1\rfloor \times \lfloor\frac{n_y-k_y}{S_y}+1\rfloor \times \lfloor\frac{n_z-k_z}{S_z}+1\rfloor$ elements of $(R\star N[r_x,r_y,r_z])_{r_xr_yr_z}$, where $r_j\in\{0,S_j,2S_j,\ldots,\lfloor\frac{n_j-k_j}{S_j}+1\rfloor\};\quad j= x,y,z$, and $\lfloor\rfloor$ represents the floor operator, telling us that we only compute elements of $C$ when the shape of $K$ is entirely contained within $N$. We call $S_x$, $S_y$, and $S_z$, the convolutional stride in $x$, $y$, and $z$, and will more compactly use $S\equiv (S_x,S_y,S_z)$ to denote the convolutional stride of a 3D convolution operation. \autoref{fig:2} shows an example of the shapes of input and output feature maps, where a $2\times 2\times 2$ filter acts on a $16\times 16\times 16$ input feature map with a convolutional stride of $S = (2,2,2)$, leading to an $8\times 8\times 8$ output grid. We can also use zero padding where we pad the outer perimeter of our input grid, $N$ with $P = (P_x,P_y,P_z)$ layers of zeros. Here, $P_j$ is a positive integer or zero and $j=x,y,z$. More generally, a $k_x\times k_y\times k_z$ kernel acting with a stride of $(S_x,S_y,S_z)$ on an $n_x\times n_y\times n_z$ input feature map with additional zero padding of $(P_x,P_y,P_z)$, will lead to an output feature map of shape $(C_x\times C_y\times C_z)$, where

\begin{align}
\label{eq:3dconv}
    C_j = \left\lfloor\frac{n_j-k_j+2P_j}{S_j}+1\right\rfloor;\quad j=x,y,z.
\end{align}
In a 3DCNN, the elements of the convolutional filters, $K_{lmn}$ are learnable parameters, so the network attempts to learn the features of interest to extract from a feature map.

\subsubsection{3D Pooling}
Pooling is an operation commonly used in convolutional neural networks to downsample a feature map while still retaining important information for classification. In our 3DCNN we include an average pooling layer (called AvgPool; see Section \ref{subsec:architecture}). Similar to a 3D convolution, the average pooling operation involves a 3D filter of a given size sliding over an input feature-map, $N$, but the average pooling operation simply computes the mean of all elements contained in the sub-block of $N$ that the filter is sliding over. The average pooling operation will thus lead to an output feature map $C$ of shape given by Equation (\ref{eq:3dconv}). Since average pooling simply computes the mean of all elements in a sub-block of $N$, there are no learnable parameters associated with average pooling.

\subsection{Data processing for the 3DCNN}

We simulate 12,380,422 electron recoil events ranging in ionization energies between \SI{0.5}{keV_{ee}} and \SI{10.5}{keV_{ee}}, 367,984 F recoil events, and 338,909 He recoils. The nuclear recoils are simulated with a uniform energy spectrum between \SI{5}{keV_{r}} and \SI{50}{keV_{r}}. While our simulated detector is sensitive to individual electrons, nuclear recoils still lose energy due to quenching, so in the absence of reconstructing the true recoil energy, we compute the ionization energy of all nuclear recoil species in terms of electron equivalent energy to allow for a direct comparison of electron rejection at a given ionization energy in our detector. Expressed in terms of electron equivalent energy, the ionization energy of a nuclear recoil $E_\text{ionization}$, is computed as

\begin{align}
    E_\text{ionization} = N_e W_e,
\end{align}
where $N_e$ is the number of electrons produced in the recoil event, and $W_e$ is the average energy per electron-ion pair. Following the approach in Ref. \cite{ghrear} we set $W_e$ to \SI{32.4}{eV}. After computing $E_\text{ionization}$, we restrict our nuclear recoil sample to $\SI{0.5}{keV_{ee}}\leq E_\text{ionization} \leq \SI{10.5}{keV_{ee}}$, leaving us with 224,299 F recoils and 93,774 He recoils.

\begin{table}[tbp]
\centering
\begin{tabular}{|c|c|c|c|c|c|}
\hline
\begin{tabular}[c]{@{}c@{}} Recoil \\ species \end{tabular} & \begin{tabular}[c]{@{}c@{}} Original energy \\ range [\SI{}{keV_{r}}] \end{tabular} & \begin{tabular}[c]{@{}c@{}} Energy \\ range [\SI{}{keV_{ee}}] \end{tabular} & \begin{tabular}[c]{@{}c@{}} $\#$ Events \\ training set \end{tabular} & \begin{tabular}[c]{@{}c@{}} $\#$ Events \\ validation set \end{tabular} & \begin{tabular}[c]{@{}c@{}} $\#$ Events\\ test set \end{tabular} \\
\hline
e  & \longdash[2]  & 0.5-10.5 & 2,947,702 & 491,429 & 8,941,291 \\
F  & 5-50 & 0.5-10.5 & 53,760 & 8,812   & 161,727 \\
He & 5-50 & 0.5-10.5 & 22,443 & 3,743 & 67,588 \\
\hline
\end{tabular}
\caption{\label{tab:train} Event samples used with the 3DCNN classifier. Events in each sample are binned in \SI{1}{keV_{ee}} wide bins centered at integer ionization energy steps between \SI{1}{keV_{ee}} and \SI{10}{keV_{ee}}. We train and evaluate the 3DCNN classifier separately for each energy bin.}
\end{table}

The charges in our simulated events are already binned with $(100\times 100\times 100)\SI{}{\um^3}$ segmentation, however because diffusion is larger than our bin size, we use a low density gas, and we do not simulate any charge amplification, we find that the charge clouds are relatively sparse at this resolution. While it is feasible to train a 3DCNN on current-generation hardware using the native $(100\times 100\times 100)\SI{}{\um^3}$ resolution of our simulation, we opt to reduce the spatial segmentation of our events to a $32\times 32\times 32$ grid of bins evenly spaced within a cube of width \SI{2.72}{cm}, leading to bin sizes of about $(850\times 850\times 850)\SI{}{\um^3}$. We do this for two reasons: First, all fluorine recoil events are entirely contained within this cube, so no recoil information is lost due to cropping, and second, the reduced spatial segmentation leads to more than a factor of 600 reduction in grid volume compared to the original $(100\times 100\times 100)\SI{}{\um^3}$ segmentation, leading to a substantial reduction in computational cost. Comparing the top and bottom rows of Figure \ref{fig:0} it appears that the structure of recoil tracks isn't significantly altered with this reduced resolution, suggesting that event classification performance for the simulated detector configuration will not be significantly hampered by our reduction of spatial segmentation. We note that uncertainties in bin placement from re-binning may slightly alter performance compared to binning the unbinned initial coordinates of 3D charge into the $32\times 32\times 32$ grid.

We use the \texttt{PyTorch} \cite{pytorch} software library for all neural network computations and store each event as a tuple containing two entries: (1) a $(32\times 32\times 32)$ voxel grid filled with binned charge that is stored as a \texttt{PyTorch} tensor data structure, and (2) an integer representing the class label of the event. We label electron recoils as 0, F recoils as 1, and He recoils as 2. Charge in each bin of the voxel grid is stored as unsigned 8 bit integers leading to an effective dynamic range of 0 to 255 electrons per voxel. No bins are saturated in any event with this dynamic range.

\subsection{Network architecture}
\label{subsec:architecture}
\begin{figure*}[htbp]
\centering 
\includegraphics[width=\linewidth]{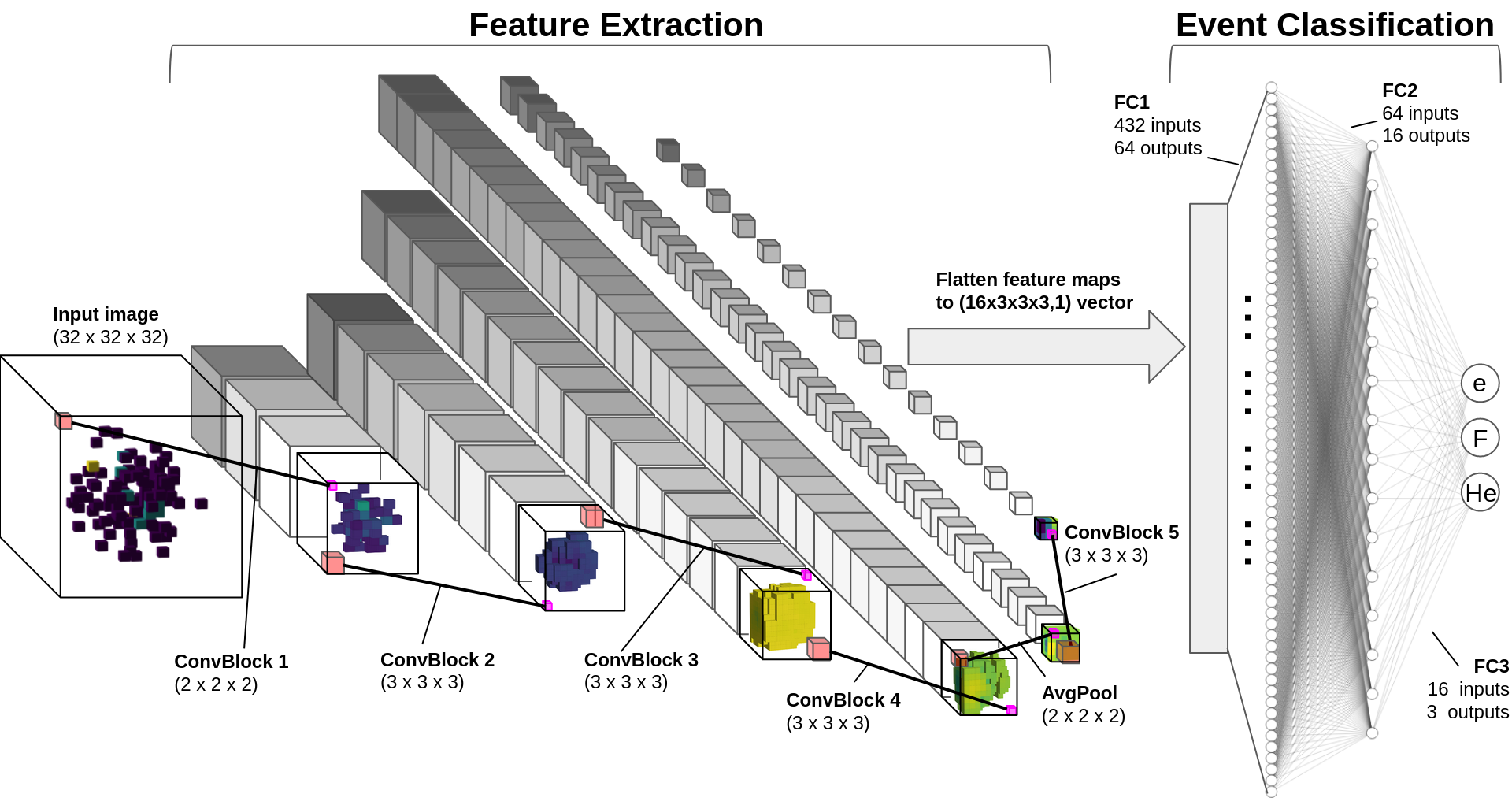}
\caption{\label{fig:cnn} Architecture of our 3DCNN classifier. Within the feature extraction portion, each cube represents a unique feature map (represented by the different shade of each map), with the size of the cubes shown approximately to scale in reference to the dimensions of the input image. We show the bin content of one such-feature map in each layer to represent the steps of a convolutional chain. Solid black lines connect larger pale red cubes, which illustrate the convolutional or pooling filters acting on a portion of the input feature map, to the smaller magenta cubes which are the outputs of the convolutional filter. The final 16 feature maps are then flattened and passed into a fully connected neural network for classification. The FCNN diagram was produced using \cite{nn}.}
\end{figure*}

\label{subsec:training}

Figure \ref{fig:cnn} and Table \ref{tab:network} together outline the network architecture of our 3DCNN. As is shown in Figure \ref{fig:cnn}, our network consists of a series of convolutional and pooling layers for feature extraction, followed by a dense fully connected neural network (FCNN) for event classification. Within the feature extraction portion of the neural network, we employ five convolutional blocks (ConvBlocks 1-5) and a pooling layer (AvgPool) to downsample the feature maps while still maintaining important features. Each of the five convolutional blocks contain the following components: (i) A 3D convolution with a convolutional filter size listed in Table \ref{tab:network}, (ii) a 3D batch normalization \cite{batch}, (iii) a scaled exponential linear unit (SELU) activation layer \cite{selu}, and (iv) a randomized dropout \cite{dropout} of 0.03 to reduce overfitting.

\begin{table}[tbp]
\centering
\setlength\tabcolsep{4.5pt}
\begin{tabular}{|c|c|c|c|c|c|c|c|}
\hline
Layer & \begin{tabular}[c]{@{}c@{}}$\#$ \\Filters \end{tabular} & Filter size & $S$ & \begin{tabular}[c]{@{}c@{}} $\#$ Learnable\\ parameters \end{tabular} & Dropout & Output shape \\
\hline
ConvBlock 1 & 4&$(2\times 2\times 2)$ &(2,2,2) & 44 & 0.03 & $(4\times 16 \times 16\times 16)$ \\
ConvBlock 2 & 8&$(3\times 3\times 3)$ &(1,1,1) & 888 & 0.03 & $(8\times 14 \times 14\times 14)$ \\
ConvBlock 3 & 16&$(3\times 3\times 3)$ &(1,1,1) & 3,504 & 0.03 & $(16\times 12 \times 12\times 12)$ \\
ConvBlock 4 & 32&$(3\times 3\times 3)$ &(1,1,1)  & 13,920 & 0.03 & $(32\times 10 \times 10\times 10)$ \\
AvgPool & 32& $(2\times 2\times 2)$ & (2,2,2) &\longdash[2] & \longdash[2] & $(32\times 5 \times 5\times 5)$ \\
ConvBlock 5 & 16& $(3\times 3\times 3)$ &(1,1,1) & 13,872 & 0.03 & $(16\times 3 \times 3\times 3)$ \\
FC1 & \longdash[2] & \longdash[2]& \longdash[2]& 27,712 & 0.05 & $(64\times 1)$ \\
FC2 & \longdash[2] & \longdash[2]& \longdash[2]& 1,040 & 0.05 & $(16\times 1)$ \\
FC3 & \longdash[2] & \longdash[2]& \longdash[2]& 51 & \longdash[2] & $(3\times 1)$ \\
\hline
\end{tabular}
\caption{\label{tab:network} More specific details of each layer shown in Figure \ref{fig:cnn}. We assume a single ($1\times 32\times 32\times 32$) image is fed into the network. The output shape column gives the shape of the output after each layer. The output of ConvBlocks 1-5 and AvgPool is a tensor of shape $(\text{D}\times \text{L}\times \text{W}\times \text{H})$, where D is the layer depth (number of feature maps), and L, W, and H are the length, width and height of each feature map, respectively. $S$ is the convolutional stride of the layer. The output shapes of the FC layers are 1 dimensional vectors for each node in a given layer. In addition to the weights and biases associated with each node in the FCNN, the entries within each convolutional filter are also learnable parameters, so we list the total number of learnable parameters associated with each layer in the network.}
\end{table}

Walking through the network architecture, we start with a ($32\times 32\times 32$) input image. The vast majority of voxels in all input images are filled with 0, so we implement a stride of 2 in ConvBlock 1 to immediately downsample this block's four output feature maps to ($16\times16\times16$) to reduce the computational overhead of training and evaluating our network. Following the chain in Figure \ref{fig:cnn}, we next perform three successive convolutional blocks (ConvBlocks 2-4) where we gradually increase the number of convolutional filters employed in each layer to produce more feature maps for classification. ConvBlocks 2-4 each use ($3\times 3\times 3$) convolutional filters with a stride of 1 and no padding, so each of these convolutions decreases the feature map dimension by two, leaving us with 32 feature maps of size ($10\times 10\times 10)$ at the end of ConvBlock4. After this, we perform average pooling. We use a ($2\times 2\times 2$) pooling filter with a stride length of 2, leaving us with 32 ($5\times 5\times 5$) feature maps that we feed into our final convolutional block (ConvBlock5) which leaves us with 16 ($3\times 3\times 3$) feature maps. Since the values composing each convolutional filter are learnable parameters, we expect that when the network is trained, there will be useful features encoded in some of these feature maps. We finally flatten these feature maps into a $((16\times 3\times 3\times 3)\times 1) = (432\times 1)$ vector, that contains each extracted feature at the end of the convolutional chain. This flattened $(432\times 1)$ feature vector is then fed into a fully connected dense neural network with two hidden layers and output class assignments corresponding to e, F, and He recoils. Each of the three fully connected layers (FC1, FC2, and FC3) use a SELU activation function and FC1 and FC2 include a random dropout of 0.05. The raw model output of each event is a $(3\times 1)$ vector, $\mathbf{z}$, with entries corresponding to each of the three class outputs (e recoil, F recoil, and He recoil). The softmax function 

\begin{align}
\label{eq:softmax}
\sigma(\mathbf{z})_i = \frac{e^{z_i}}{\sum_{j=1}^3e^{z_j}}.
\end{align}
is applied to $\mathbf{z}$ to map the class outputs $z_i\in\mathbf{z}$ to class probabilities. We henceforth label $\sigma(\mathbf{z})_1$, $\sigma(\mathbf{z})_2$, and $\sigma(\mathbf{z})_3$, as $\rm p_e$, $\rm p_F$, and $\rm p_{He}$, which represent the model-predicted class probabilities of e, F, and He recoils, respectively.

\subsection{Training the network}

We first shuffle the order of all events and then split the data into distinct training, validation, and test sets with 3,023,905 events in our training sample, 503,984 events in our validation sample, and 9,172,513 events in our test sample. We set the test sample aside and implement the following procedure to train our model:
\begin{enumerate}
    \item Form a \texttt{PyTorch} tensor of shape $(256\times 32\times 32\times 32)$, which is a \textit{minibatch} consisting of 256 randomly selected voxel images from the training sample.
    \item Feed the minibatch and the corresponding truth label of each image of the minibatch into the 3DCNN.
    \item Use \texttt{PyTorch}'s built in CrossEntropy loss function to compute the loss of the batch. We wish to minimize this loss. We use an Adam \cite{adam} optimizer with a learning rate of 0.0002.
    \item Update model weights using backpropagation \cite{backprop}.
    \item Repeat steps 1-4 until we've run through all events in the training set. This is called a training epoch.
    \item At the end of each training epoch, repeat steps 1-3 for the validation sample. We do not implement step 4 as we don't wish to train the 3DCNN on the validation set. Compute the sum of the losses of each minibatch of the validation sample.
    \item If the summed losses over the validation set minibatches are less than in the previous epoch, we treat this as the model learning and save all model weights.
    \item Implement early stopping \cite{estop} where steps 1-7 are repeated until the total validation loss doesn't decrease at all over 10 successive epochs.
    \item The model state corresponding to the epoch with the lowest validation loss is our trained model.
    
\end{enumerate}
We were able to train the 3DCNN with an Nvidia GeForce RTX 2070 consumer-grade laptop GPU in less than 6 hours. 
\begin{figure}[htbp]
\centering 
\includegraphics[width=\textwidth]{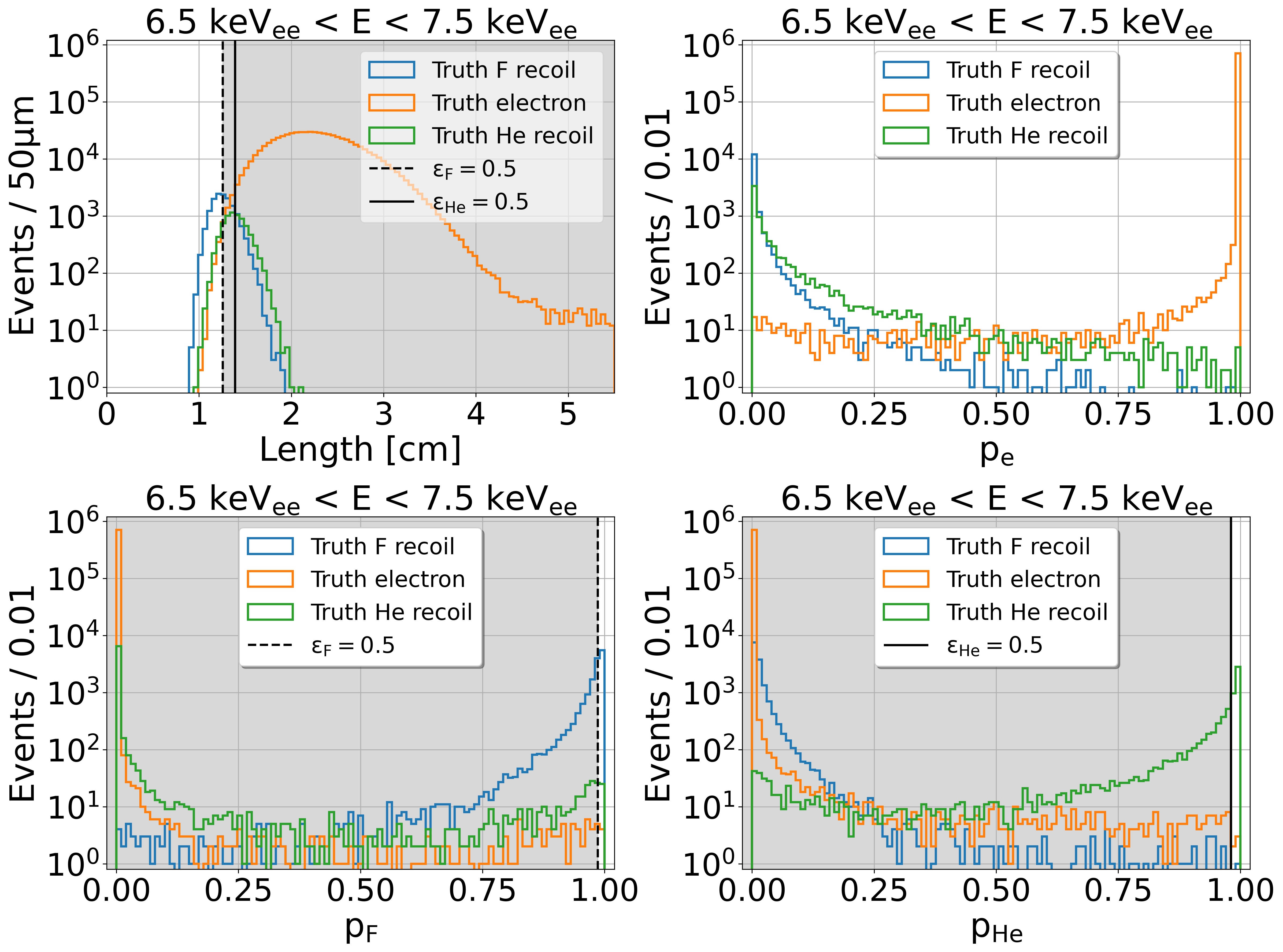}
\caption{\label{fig:3} Comparison of traditional (track length) and 3DCNN-based ($\rm p_e$, $\rm p_F$ and $\rm p_{He}$) discriminants for events in the test sample satisfying $\SI{6.5}{keV_{ee}} < E_\text{ionization} < \SI{7.5}{keV_{ee}}$. The shaded regions in the panels indicate events that are rejected after selections corresponding to $50\%$ F recoil efficiency (dashed black vertical line) and/or $50\%$ He recoil efficiency (solid black vertical line).}
\end{figure}
\subsection{Interpreting model output}
\label{subsec:interp}
We use the 3DCNN output variables $\rm p_e$, $\rm p_F$, and $\rm p_{He}$ as event classification discriminants. Figure \ref{fig:3} shows a comparison of the distributions of track length, $\rm p_{e}$, $\rm p_F$, and $\rm p_{He}$ for all true F recoils (blue), all true He recoils (green), and all true electrons (orange) in the test sample of data. Since $\rm p_{e}$, $\rm p_F$, and $\rm p_{He}$ represent class probabilities, for each event, $\rm p_{e}+\rm p_{F}+\rm p_{He} = 1$, meaning if our 3DCNN is a good classifier, we should then expect that each class probability peaks strongly toward 1 for its corresponding class, and strongly toward 0 for the other two classes. This expected behavior is observed in Figure \ref{fig:3}, suggesting that $\rm p_{e}$, $\rm p_F$, and $\rm p_{He}$ are strong event classification discriminants.

In addition to rejecting electron backgrounds, these discriminants can also be used to distinguish nuclear recoil signal candidates. To quantify classification performance, we define the signal efficiency as

\begin{align}
    \varepsilon_s = \frac{N_s'}{N_s},
\end{align} 
where $N_s$ is the total number of signal events in the sample and $N_s'$ is the number of remaining signal events after making a selection. Now, suppose we want to quantify the signal purity of F recoils at an F recoil efficiency of $\varepsilon_F = 0.5$. We can do this by determining $\rm p_F'$, the median $\rm p_F$ of true F recoils. Then, we can determine the number of F, He, and e recoils where $\rm p_F > p_F'$, and compute the F recoil purity as $N_\text{F}'/(N_\text{e}'+N_\text{F}'+N_\text{He}')$. The bottom left panel of Figure \ref{fig:3} shows an example of these selections for the set of recoil events between \SI{6.5}{keV_{ee}} and \SI{7.5}{keV_{ee}}, where the black vertical dashed line corresponds to $\rm p_F'$. We can perform an analogous procedure with $\rm p_{He}$ to compute the He recoil at a fixed He recoil efficiency.

In certain instances it may be advantageous to treat any nuclear recoil species as signal and electrons as background. In these cases we could treat $\rm p_e$ as a binary classification variable between electrons versus not electrons (i.e. signal nuclear recoils for this ternary classification model). For example, if we wanted to calculate the nuclear recoil purity at a nuclear recoil efficiency of $\varepsilon_\mathcal{R} = 0.5$, we could determine, $\rm p_e'$, the median value of $\rm p_e$ for all He or F recoils. We would then compute ($N_\text{F}'+N_\text{He}')/(N_\text{e}'+N_\text{F}'+N_\text{He}')$ where $N_\text{F}'$, $N_\text{He}'$, and $N_\text{e}'$ represent the number of events of each of these three classes after applying a selection of $\rm p_e < \rm p_e'$.

Since signal purity depends on the relative composition of recoil constituents, we opt to mostly use the notion of a rejection factor, $R$, to quantify background rejection performance. We define the sample size-independent \textit{electron} rejection factor $R_\text{e}$ as the ratio of the total number of electrons in a sample $N_\text{e}$ to the number of electrons remaining after a selection, $N_\text{e}'$

\begin{align}
\label{eq:erej}
    R_\text{e} = \frac{N_\text{e}}{N'_\text{e}}.
\end{align}
For the \SI{7}{keV_{ee}} sample shown in Figure \ref{fig:3}, when using length as the event classification variable we find rejection factors of $135\pm 2$ and $870\pm 30$ for $\varepsilon_\text{He} = 0.5$ and $\varepsilon_\text{F} = 0.5$, respectively. Using 3DCNN output probability as our event classification variable, we find that only five electrons remain both for He recoils at $\varepsilon_\text{He} = 0.5$ and F recoils at $\varepsilon_\text{F} = 0.5$, leading to an electron rejection factor of $141,400\pm 63,000$ for both of these cases, a more than 3 order of magnitude improvement over the traditional track length discriminant for He recoils.

\section{Shallow learning classifiers}
\label{sec:erej}
Here we introduce the nine event-shape discriminants used in Ref.~\cite{ghrear} and train both a BDT (Section \ref{subsec:bdt}) and an FCNN (Section \ref{subsec:nn}) to combine these observables into multivariate classification discriminants.

\subsection{Defining electron rejection discriminants}

Each of the following variables are computed using our simulated TPC's native $(100\times 100\times 100)\SI{}{\um^3}$ resolution, as opposed to the reduced $(850\times 850\times 850)\SI{}{\um^3}$ segmentation used for the 3DCNN classifier. 

\begin{enumerate}
\item Length along the principal axis (LAPA): We use a singular value decomposition (SVD) to identify the principal axis of the 3D track, and then take the difference between the maximum and minimum of the track's ionization distribution coordinates projected onto this principal axis.
\item Standard deviation of charge distribution (SDCD): The standard deviation of the 3D position vectors of all charges in the event.
\item Maximum charge density of the event ($\rho_\text{Max}$): This discriminant is optimized by varying the bin width of the charge in cubic voxels and recording the maximum amount of charge in a voxel. $\rho_\text{Max}$ was separately optimized to maximize electron rejection for weakly-directional \SI{7}{keV_{ee}} F recoils and for directional \SI{12}{keV_{ee}} He recoils as described in Ref. \cite{ghrear}.
\item Charge uniformity (QUnif): The standard deviation of the distribution of mean distances between each charge and all other charges in a recoil event.
\item Cylindrical thickness (CylThick): Sum of the squared transverse distances from each charge to the principal axis of the track.
\item Number of clusters (NClust): Number of clusters determined by the \texttt{DBSCAN} \cite{dbscan} clustering algorithm. Two of the input parameters to this algorithm were adjusted to find the pair that optimizes electron rejection both for weakly-directional \SI{7}{keV_{ee}} F recoils and for directional \SI{12}{keV_{ee}} He recoils as described in Ref. \cite{ghrear}.
\item Clustering threshold (CThres): The threshold that the fraction of total event charge in the largest cluster of an event must be above. Clustering is performed \texttt{DBSCAN} and the threshold value is optimized separately to maximize electron rejection for weakly-directional \SI{7}{keV_{ee}} F recoils and for directional \SI{12}{keV_{ee}} He recoils.
\end{enumerate}
Given the separate directional and weakly-directional optimizations for the $\rho_\text{Max}$ and ClustThres observables, we have a total of nine discriminants to work with.

\subsection{Boosted decision tree}
\label{subsec:bdt}
We use the \texttt{XGBClassifier} model in the \texttt{XGBoost} \cite{xgboost} software library as our BDT classifier for recoil classification. For this BDT analysis, we first randomize the order of our data and then partition the data into a $23.8\%$ / $4.0\%$ / $72.2\%$ training/validation/testing sample split. The validation data sample is not used for the BDT analysis; however we still opt to create a validation set so that we have an identical testing set to what's later used in the FCNN analysis. For each event, we compute the nine discriminants and store the results as a $(9\times 1)$ vector to feed into the BDT classifier. We assign class labels of 0 for e recoils, 1 for F recoils, and 2 for He recoils.

\begin{table}[tbp]
\centering
\begin{tabular}{|c|c|c|c|c|c|c|c|c|}
\hline
$\rho_\text{Max,F}$ & SDCD & CThres$_\text{F}$ & $\rho_\text{Max,He}$ & CThres$_\text{He}$ & CylThick & QUnif & LAPA & NClust \\
\hline
0.437 & 0.231 & 0.169 & 0.111 & 0.032 & 0.009 & 0.007  & 0.003 & 0.002 \\
\hline
\end{tabular}
\caption{\label{tab:features} Normalized feature importance of the nine discriminants in the trained BDT.}
\end{table}

Given the relatively quick training and evaluation time of \texttt{XGBoost} compared to our 3DCNN classifier, we perform a course, two-step partial grid search to optimize some of the hyperparameters in the \texttt{XGBClassifier} model. We make the \textit{a priori} decision to choose the set of hyperparameters $(N_\text{trees}, D_\text{tree}, \ell)$ that maximizes the electron rejection factor $R_\text{e}$ for \SI{7}{keV_{ee}} F recoils at $\varepsilon_\text{F} = 0.5$, where $N_\text{trees}$, $D_\text{tree}$, and $\ell$ represent the number of trees, the maximum depth of a tree, and the learning rate of the model, respectively. The first step of our grid search is to fix $\ell$ at 0.2, and then train and evaluate our BDT using all 16 combinations of $N_\text{trees}\in\{50,100,200,500\}$ and $D_\text{tree}\in \{2,3,5,7\}$. We find that $N_\text{trees} = 200$ with $D_\text{tree} = 3$ to be the best combination at $\ell = 0.2$, leading to $R_\text{e} = 101,000\pm 38,000$. Since this combination of $N_\text{trees}$ and $D_\text{tree}$ does not sit at the boundary of our grid, we keep this combination and move on to our next step of optimizing $\ell$. In our second step of the grid search, we train our classifier using $\ell\in\{0.1,0.15,0.2,0.25,0.3\}$ and find our initial choice of $\ell = 0.2$ to be the best, so we freeze our optimal set of hyperparameters for our BDT classifier at $(N_\text{trees} = 200,D_\text{tree} = 3,\ell=0.2)$. 

We trained our BDT model using 53,959 F recoils, 22,265 He recoils and 2,992,534 e recoils in our training sample\footnote{Data for the shallow learning classifiers uses the same simulation as the 3DCNN, but was processed independently, leading to small differences in the training sample statistics compared to Table \ref{tab:train}. We ensured the number of each recoil species in the \textit{test} sample here matches exactly with Table \ref{tab:train}. Furthermore, when binning in integer steps of electron equivalent energy, the number of each recoil species in the test sample here is identical to the number of each recoil species in the 3DCNN test sample within each energy bin.}. Similar to the 3DCNN classifier, the BDT outputs ternary class probabilities, $\rm p_{e,BDT}$, $\rm p_{F,BDT}$, and $\rm p_{He,BDT}$, allowing for the same model output interpretation as the 3DCNN (Section \ref{subsec:interp}).

A benefit of using a BDT classifier is the ability to track the decisions made throughout all branches in the trained classifier model. Table \ref{tab:features} shows the normalized feature importance of each of the nine discriminants. Here we define the feature importance for a particular observable to be the sum of the information gain \cite{gain} of all splits where the observable is used, normalized so that the sum of the feature importance of all nine observables is 1. The two observables with the highest feature importance, $\rho_\text{Max,F}$ and SDCD, are the same two discriminants that perform best in terms of electron rejection factor versus $E_\text{ionization}$ at $50\%$ F recoil efficiency in Ref.~\cite{ghrear}.

\subsection{Fully connected neural network}
\label{subsec:nn}

We next train and evaluate the performance of an FCNN using identical training, validation, and test sets to those generated for the BDT classifier to derive a fair comparison between the two classifiers. Our FCNN consists of an input layer with 9 nodes representing each of the electron rejection observables we've defined, 3 hidden layers, and an output layer with 3 nodes that represent our three output classes (see Table \ref{tab:nn}). We use a ReLU activation with batch normalization after the input and hidden layers, and apply a randomized dropout of 0.05 in all layers except the output layer. We train this FCNN classifier by forming minibatches of 512 events, where each event consists of a ($9\times 1$) \texttt{PyTorch} tensor of normalized $z$-scores of each of the nine observables and the associated class label. We then follow the same training procedure (steps 3-9 shown in Section \ref{subsec:training}) with the same early stopping trigger that we used for the 3DCNN classifier training, except we use a learning rate of 0.001.
\begin{table}[tbp]
\centering
\begin{tabular}{|c|c|c|c|c|c|c|c|c|}
\hline
Layer & Input shape & Output shape & Dropout \\
\hline
Input & ($9\times 1$) & ($32\times 1$) & 0.05 \\
Hidden 1 & ($32\times 1$) & ($64\times 1$) & 0.05 \\
Hidden 2 & ($64\times 1$) & ($128\times 1$) & 0.05 \\
Hidden 3 & ($128\times 1$) & ($32\times 1$) & 0.05 \\
Output & ($32\times 1$) & ($3\times 1$) & \longdash[2] \\
\hline
\end{tabular}
\caption{\label{tab:nn} FCNN classifier architecture. As mentioned in the text, the input vectors are formed with normalized $z$-scores of each of the nine electron rejection observables. We use a ReLU activation function and a randomized dropout of 0.05 at the input and each of the hidden layers. When test sample events are passed through the trained network, a softmax function is applied to the model output to give class probabilities.}
\end{table}

We performed a moderate amount of hand-tuning of the hyperparameters for the FCNN, rather than a full hyperparameter search, and find that in general the FCNN gives better electron rejection performance than the BDT classifier, especially for He recoils. Similar to the 3DCNN, the output of each event passed through the FCNN classifier is a $(3\times 1)$ vector containing outputs associated with each of the three recoil species, so we apply a softmax to this vector to obtain class probabilities $\rm p_{e,FCNN}$, $\rm p_{F,FCNN}$, and $\rm p_{He,FCNN}$ that may be interpreted analogously to the output class probabilities given by the 3DCNN and BDT.

\section{Event identification performance results}
\label{sec:results}
We now have a set of class probabilities from the 3DCNN, BDT, and FCNN classifiers that can be used to quantify and compare event selection performance. We also include the multivariate observable from Ref. \cite{ghrear} in our comparisons where available.

\begin{figure*}[htbp]
\centering 
\includegraphics[width=\linewidth]{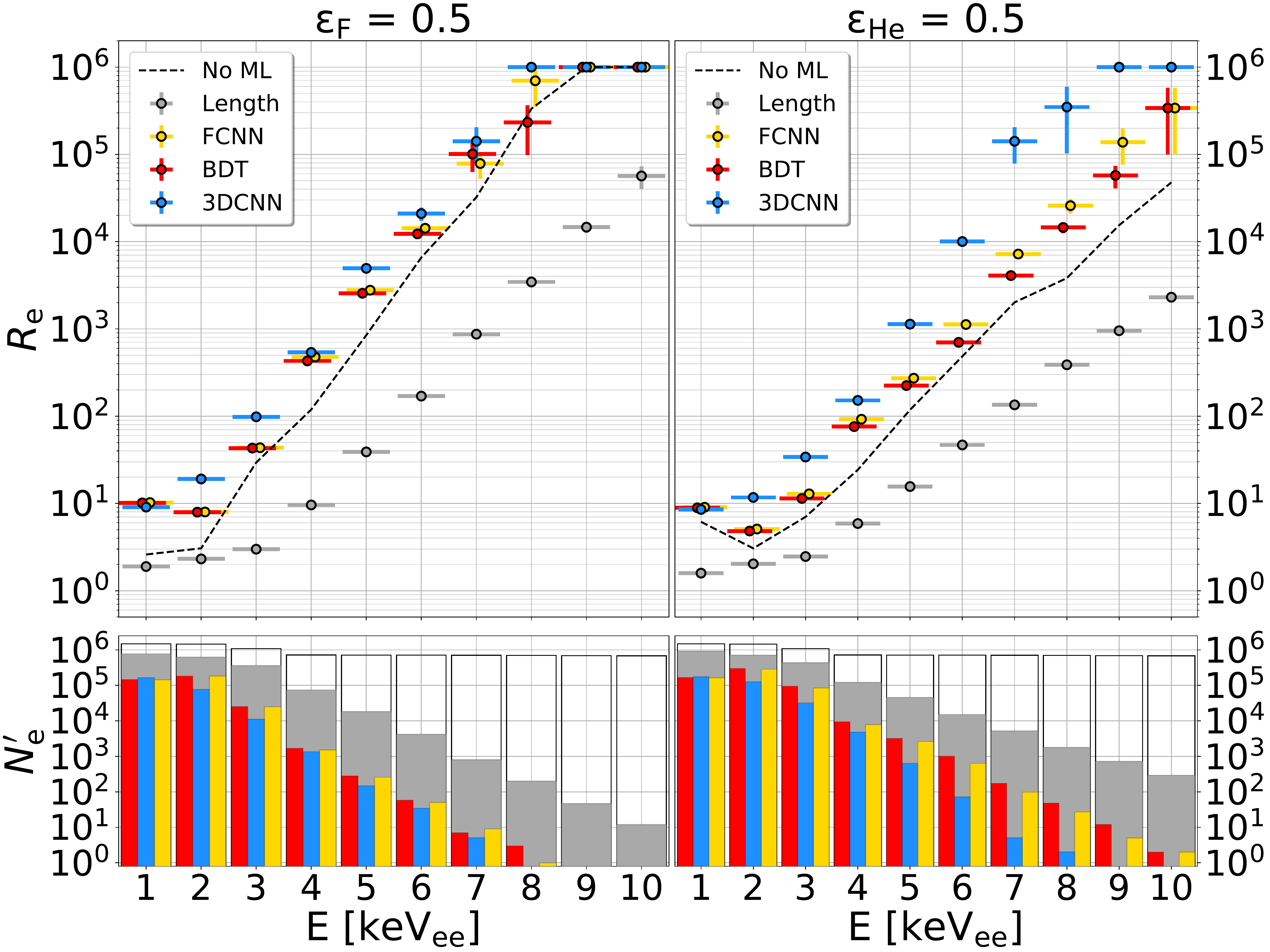}
\caption{\label{fig:6} Comparison of electron-rejection performance of all discriminants versus electron-equivalent energy at $50\%$ fluorine recoil efficiency (left) and $50\%$ helium recoil efficiency (right). The top and bottom row plots share a common horizontal axis. The black dashed lines show the combined observable introduced in Ref. \cite{ghrear}. Data within a given energy bin in the $R_\text{e}$ plots are staggered slightly to separate overlapping points. The bottom bar plots show the number of remaining electrons after a given recoil selection in each integer energy bin. The transparent bars indicate the total number of electron recoils in each energy bin.}
\end{figure*}

\begin{figure*}[htbp]
\centering 
\includegraphics[width=\linewidth]{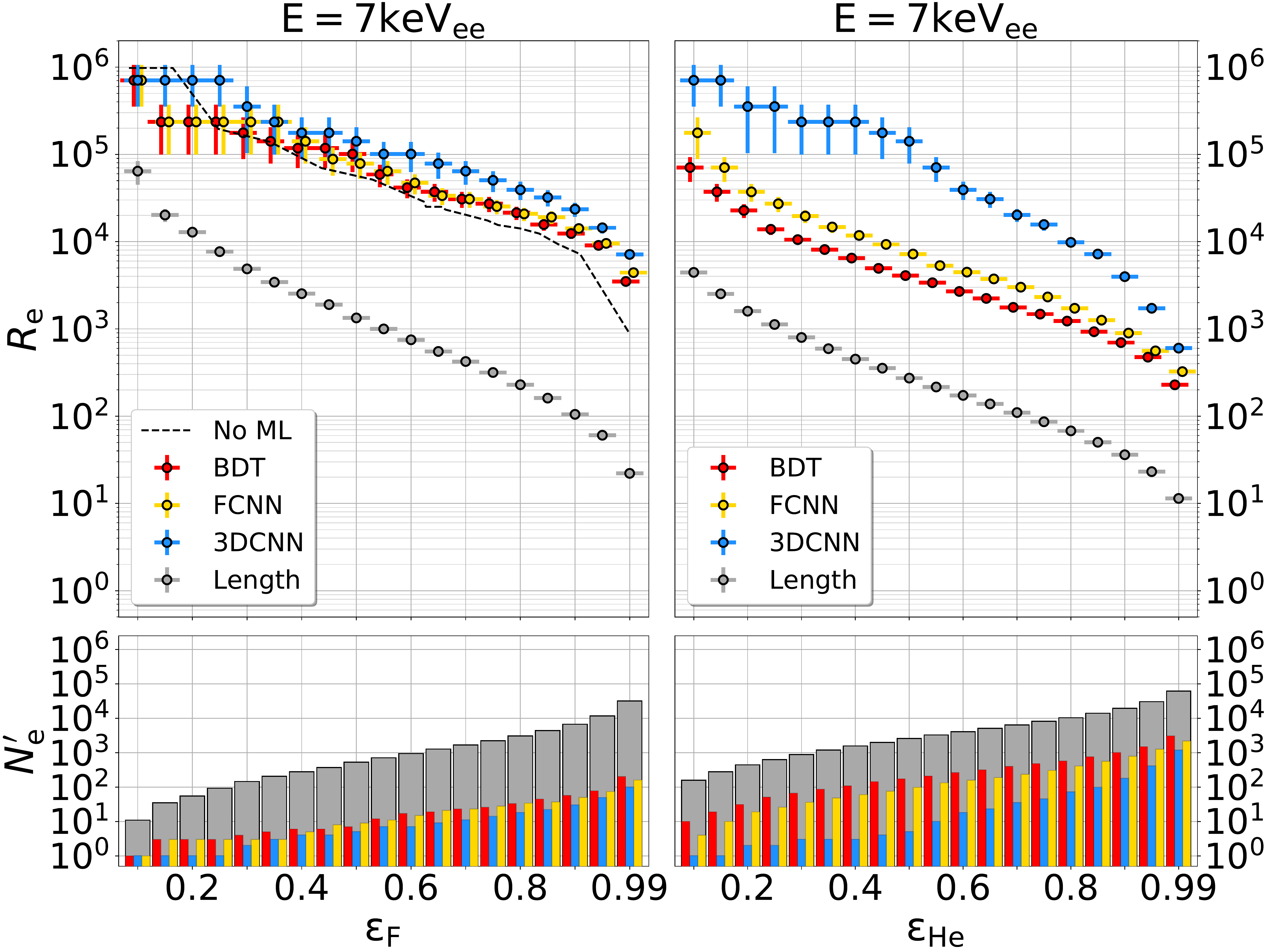}
\caption{\label{fig:5} Comparison of electron-rejection performance of all discriminants versus nuclear recoil efficiency for \SI{7}{keV_{ee}} fluorine recoils (left) and \SI{7}{keV_{ee}} helium recoils (right). The top and bottom row plots share a common horizontal axis. The black dashed line in the upper left hand plot shows the combined observable from Ref. \cite{ghrear}. Data points at a given recoil efficiency in the $R_\text{e}$ plots are staggered slightly to separate overlapping points. The bottom bar plots show the number of remaining electrons after a given recoil selection in each efficiency bin.}
\end{figure*}

For a future directional DM detector, it is useful to consider electron background rejection both as a function of ionization energy at fixed nuclear recoil selection efficiency and as a function of nuclear recoil efficiency at fixed energies. Comparing electron rejection performance versus energy provides insight toward the energy threshold required for near background-free operation at a given exposure, while comparing versus nuclear recoil efficiency allows us to compare detector exposure required for a fixed signal yield. Figures \ref{fig:6} and \ref{fig:5} show the results for both of these approaches.

Figure \ref{fig:6} shows $R_\text{e}$ performance versus energy at a fixed $50\%$ nuclear recoil efficiency for F recoils (left) and He recoils (right). In cases where all electrons are rejected, we plot $R_\text{e}$ as $1\times 10^6$ with zero uncertainty. The black dashed line labeled ``No ML'' is the combined observable from Ref. \cite{ghrear}. For the three machine learning discriminants and the traditional track length discriminant, we apply the following procedure to determine $R_\text{e}$ in each energy bin:
\begin{enumerate}
    \item Determine the values, $\rm p_{\mathcal{R}}'$, $\rm p_{\mathcal{R},BDT}'$, $\rm p_{\mathcal{R},FCNN}'$, and track length, $L_\mathcal{R}'$, that correspond to a 50$\%$ efficiency of recoil $\mathcal{R}$. In the left panel of Figure \ref{fig:6}, $\mathcal{R}=\text{F}$, and in the right panel, $\mathcal{R}=\text{He}$.
    \item $N_\text{e}$ is the total number of electrons within the energy bin and $N_\text{e}'$ for the 3DCNN, BDT, FCNN, and track length classifiers, is the number of electrons in the energy bin satisfying selections of $\rm p_{\mathcal{R}} > p_{\mathcal{R}}'$, $\rm p_{\mathcal{R},BDT} > p_{\mathcal{R},BDT}'$, $\rm p_{\mathcal{R},FCNN} > p_{\mathcal{R},FCNN}'$, and $L_{\mathcal{R}} < L_{\mathcal{R}}'$, respectively.
    \item $R_\text{e}$ is computed for the 3DCNN, BDT, FCNN, and track length classifiers with Eq. (\ref{eq:erej}) using the $N_\text{e}'$ associated with the chosen classifier. 
\end{enumerate}
In general, we find that the 3DCNN outperforms all other methods of electron background rejection, especially for He recoils. The CYGNUS collaboration mentions in their feasibility study that electron rejection will effectively determine their energy threshold \cite{cygnus}, and argue that with a 6 year exposure and flat electron background energy spectrum of \SI{1e4}{keV_{ee}^{-1} year^{-1}}, they will be essentially background-free at energies corresponding to an electron rejection factor in excess of $6\times 10^4$. With $\varepsilon_\text{F} = 0.5$, we find $R_\text{e}>6\times 10^4$ starting around $\SI{10}{keV_{ee}}$ using length as a discriminant, compared to somewhere between $\SI{6}{keV_{ee}}$ and $\SI{7}{keV_{ee}}$ with the 3DCNN, meaning the use of the 3DCNN for electron rejection lowers the ``background-free'' energy threshold by more than \SI{3}{keV_{ee}} over using length. At $\varepsilon_\text{He} = 0.5$ Ref. \cite{ghrear} found that $R_\text{e}$ doesn't exceed $6\times 10^4$ until around \SI{14}{keV_{ee}} using length as a discriminant, while with the 3DCNN classifier it exceeds $6\times 10^4$ at \SI{7}{keV_{ee}}. We note here that given the logarithmic scale shown in Figure \ref{fig:6}, electron rejection performance drops off exponentially with energy, so a factor of two reduction of the ``background-free'' energy threshold for He recoils when using the 3DCNN classifier over using length is a significant improvement. The otherwise state-of-the-art FCNN and BDT combinations of observables have ``background-free'' thresholds of around \SI{9}{keV_{ee}} for He recoils so the 3DCNN reduces this threshold by around \SI{2}{keV_{ee}}. Even a seemingly small reduction in energy threshold has a significant impact on DM reach, given the that expected DM recoil energy spectrum falls steeply with energy.

Figure \ref{fig:5} shows $R_\text{e}$ and $N_\text{e}'$ after selections at various F recoil efficiencies (left) and He recoil efficiencies (right) for these same discriminants. We choose to evaluate $R_\text{e}$ and $N_\text{e}'$ at a fixed electron-equivalent energy range of $\SI{6.5}{keV_{ee}}< E < \SI{7.5}{keV_{ee}}$ because this is the lowest energy range exceeding the CYGNUS background-free criteria of $R_\text{e} > 6\times10^4$ at both $\varepsilon_\text{F} = 0.5$ and $\varepsilon_\text{He} = 0.5$. We use the same procedure to compute $R_\text{e}$ and $N_\text{e}'$ as in Figure \ref{fig:6}, except now the values of $\rm p_{\mathcal{R}}'$, $\rm p_{\mathcal{R},BDT}'$, $\rm p_{\mathcal{R},FCNN}'$, and $L'$ are separately computed for each $\varepsilon_{\mathcal{R}}$. Since the $\rho_\text{Max,F}$, $\text{CThres}_\text{F}$, and NClust variables were optimized specifically for \SI{7}{keV_{ee}} F recoils, we expect this \SI{7}{keV_{ee}} bin will produce the best relative $R_\text{e}$ of the shallow learning classifiers compared to the 3DCNN. We note here that the combined observable from Ref. \cite{ghrear} was computed at \SI{12}{keV_{ee}} for He recoils versus $\varepsilon_\text{He}$. Given that the 3DCNN rejects all but two electrons down to $\SI{8}{keV_{ee}}$ at $\varepsilon_\text{He}=0.5$ (Figure \ref{fig:6}), we expect the 3DCNN to reject essentially all \SI{12}{keV_{ee}} electron recoils, so we do not investigate the \SI{12}{keV_{ee}} scenarios discussed in \cite{ghrear}.

\begin{figure*}[htbp]
\centering 
\includegraphics[width=\linewidth]{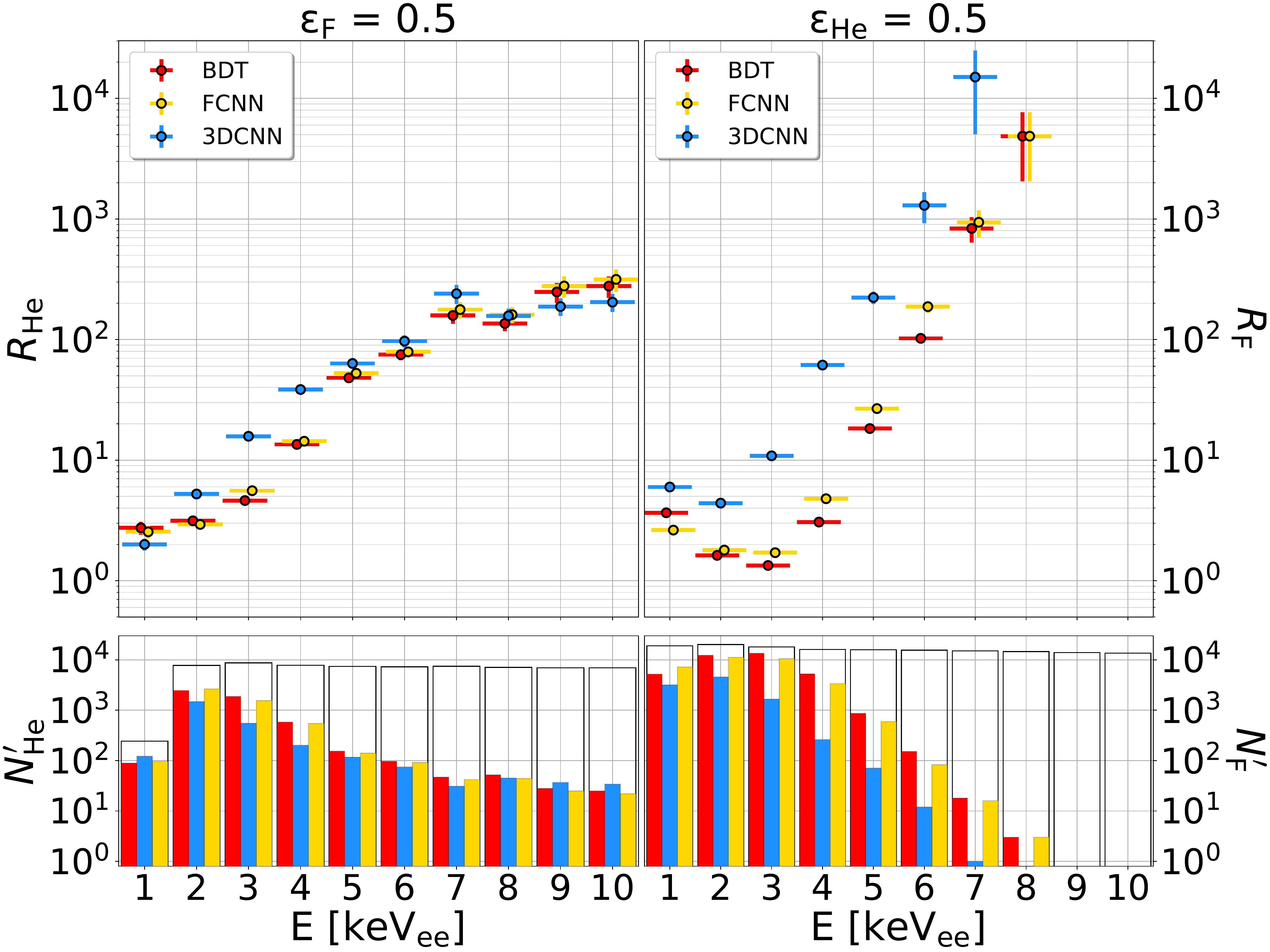}
\caption{\label{fig:7} Comparison of He and F recoil rejection at fixed F and He signal efficiencies. $R_\text{He}$ and $R_\text{F}$ are rejection factors for He and F recoils, respectively and are both defined analogously to $R_\text{e}$. The bins in the upper right plot with no data points plotted have all F recoils rejected.}
\end{figure*}

Comparing each multivariate classifier constructed from combining the nine event-shape variables, we find that introducing machine learning for multivariate analysis leads to an improvement in electron rejection above $\varepsilon_\text{F}\approx 0.2$. We find further improvement with the use of the 3DCNN, where it improves $R_\text{e}$ by more than 2 orders of magnitude over selections on length at higher F recoil efficiencies. For He recoils, we find that the 3DCNN classifier significantly outperforms all other models, often leading to an order of magnitude increase in $R_\text{e}$ compared to the BDT and FCNN combinations of the nine electron rejection observables. Furthermore, we find that selecting on 3DCNN output probability at $85\%$ He recoil efficiency leads to a higher $R_\text{e}$ than selecting on length at $10\%$ He recoil efficiency, meaning that usage of the 3DCNN classifier could allow an experiment to run with a factor of 8.5 smaller exposure and obtain the same He recoil signal, at an improved background level over using track length to classify events.

An additional benefit of training the machine learning models as ternary classifiers is the ability to use the three output class probabilities to not only reject electron backgrounds, but also classify nuclear recoil species. Figure \ref{fig:7} summarizes the nuclear recoil classification ability of the three machine learning classifiers. Above \SI{6}{keV_{ee}} all three classifiers maintain 50$\%$ F recoil efficiency while rejecting greater than $99\%$ of He recoils. Above \SI{5}{keV_{ee}}, each classifier starts to do a better job selecting pure samples of He recoils than F recoils, especially the 3DCNN classifier. Selecting pure samples of He recoils is desirable, as low energy He recoils tend to have better angular resolution than F recoils of the same energy, so the ability to select a background-free, almost entirely pure sample of He recoils down to \SI{7}{keV_{ee}} at $\varepsilon_\text{He}=0.5$ is significant. To quantify this, there are 3,721 He recoils, 1 F recoil, and 5 e recoils in our sample that satisfy $\rm p_{He} > p_{He}'$, where $\rm p_{He}'$ is the 3DCNN He class probability corresponding to $\varepsilon_\text{He}=0.5$ in the \SI{7}{keV_{ee}} energy bin. This means we have a He recoil purity in excess of $99.8\%$ for \SI{7}{keV_{ee}} He recoils at $\varepsilon_\text{He}=0.5$. Though the fraction of each event type present in our simulation doesn't model a physically expected event composition, these numbers suggest that use of the 3DCNN will lead to a high He recoil purity for \SI{7}{keV_{ee}} He recoils at $\varepsilon_\text{He}=0.5$ in a directional DM experiment.

\section{Conclusion}
We have introduced a deep learning approach to event selection in 3D recoil-imaging TPCs and compared its performance to other state-of-the-art machine learning-based multivariate classifiers. Training a 3D convolutional neural network on the 3D charge distributions of recoil tracks, we found a significant improvement in electron rejection performance between \SI{0.5}{keV_{ee}} and \SI{10.5}{keV_{ee}} compared to shallow learning methods that form multivariate classification variables from predefined event-shape observables. Notably, the improved performance of the 3DCNN holds when using input data with a reduced spatial segmentation of $(850\times 850\times 850)$\SI{}{\um^3} compared to $(100\times 100\times 100)$\SI{}{\um^3} segmentation used to compute the observables fed into the shallow learning classifiers. We find that the 3DCNN classifier outperforms other methods by a larger margin for He recoils than for F recoils indicating that the nine event-shape observables used in the BDT and FCNN classifiers are better suited for electron rejection in F recoil samples than in He recoil samples. Since the 3DCNN classifier decides which features for event classification are most useful on its own, using a classifier like this is advantageous, as it seems to have found more useful sets of features to classify nuclear recoils (especially He recoils) than our predefined observables. Defining the background-free energy threshold of our simulated detector to be the lowest energy corresponding to $R_\text{e}\geq 6\times 10^4$, we find our threshold to be between \SI{6}{keV_{ee}} and \SI{7}{keV_{ee}} for both F and He recoils at $50\%$ nuclear recoil efficiency when using the 3DCNN output probability to select for events. Using this 3DCNN to classify events effectively reduces the energy threshold of our simulated detector by over $30\%$ for F recoils and around $50\%$ for He recoils compared to classification using track length; a significant improvement over traditional keV-scale recoil identification techniques considering the steeply falling energy spectrum expected for WIMP recoils. The 3DCNN classifier is also able to assign classification probabilities to multiple recoil species simultaneously, making it a flexible approach for event classification that should be of general interest for directional DM searches.
\acknowledgments
This work was supported by the U.S. Department of Energy (DOE) via Award Number DE-SC0010504.

\bibliography{nn.bib}

\begin{thebibliography}{29}
\providecommand{\natexlab}[1]{#1}
\providecommand{\url}[1]{\texttt{#1}}
\expandafter\ifx\csname urlstyle\endcsname\relax
  \providecommand{\doi}[1]{doi: #1}\else
  \providecommand{\doi}{doi: \begingroup \urlstyle{rm}\Url}\fi

\bibitem[Spergel(1988)]{spergel}
David~N. Spergel.
\newblock Motion of the earth and the detection of weakly interacting massive
  particles.
\newblock \emph{Phys. Rev. D}, 37:\penalty0 1353--1355, Mar 1988.
\newblock \doi{10.1103/PhysRevD.37.1353}.
\newblock URL \url{https://link.aps.org/doi/10.1103/PhysRevD.37.1353}.

\bibitem[Mayet et~al.(2016)]{Mayet:2016zxu}
F.~Mayet et~al.
\newblock {A review of the discovery reach of directional Dark Matter
  detection}.
\newblock \emph{Phys. Rept.}, 627:\penalty0 1--49, 2016.
\newblock \doi{10.1016/j.physrep.2016.02.007}.

\bibitem[Vahsen et~al.(2021)Vahsen, O'Hare, and Loomba]{Vahsen:2021gnb}
Sven~E. Vahsen, Ciaran A.~J. O'Hare, and Dinesh Loomba.
\newblock {Directional Recoil Detection}.
\newblock \emph{Ann. Rev. Nucl. Part. Sci.}, 71:\penalty0 189--224, 2021.
\newblock \doi{10.1146/annurev-nucl-020821-035016}.

\bibitem[O'Hare et~al.(2022)]{OHare:2022jnx}
C.~A.~J. O'Hare et~al.
\newblock {Recoil imaging for dark matter, neutrinos, and physics beyond the
  Standard Model}.
\newblock In \emph{{2022 Snowmass Summer Study}}, 3 2022.

\bibitem[Vahsen et~al.(2020)]{cygnus}
S.~E. Vahsen et~al.
\newblock {CYGNUS: Feasibility of a nuclear recoil observatory with directional
  sensitivity to dark matter and neutrinos}.
\newblock 8 2020.

\bibitem[Breiman et~al.(1984)Breiman, J.H., R.A., and Stone]{bdt}
Friedman Breiman, L., Olshen J.H., R.A., and C.J. Stone.
\newblock \emph{Classification And Regression Trees}.
\newblock Routledge, 1st edition, 1984.
\newblock \doi{https://doi.org/10.1201/9781315139470}.

\bibitem[Billard et~al.(2012)Billard, Mayet, and Santos]{billard}
J~Billard, F~Mayet, and D~Santos.
\newblock Low energy electron/recoil discrimination for directional dark matter
  detection.
\newblock \emph{Journal of Cosmology and Astroparticle Physics}, 2012\penalty0
  (07):\penalty0 020–020, Jul 2012.
\newblock ISSN 1475-7516.
\newblock \doi{10.1088/1475-7516/2012/07/020}.
\newblock URL \url{http://dx.doi.org/10.1088/1475-7516/2012/07/020}.

\bibitem[Ghrear et~al.(2021)Ghrear, Vahsen, and Deaconu]{ghrear}
Majd Ghrear, Sven~E. Vahsen, and Cosmin Deaconu.
\newblock {Observables for recoil identification in high-definition Gas Time
  Projection Chambers}.
\newblock \emph{JCAP}, 10:\penalty0 005, 2021.
\newblock \doi{10.1088/1475-7516/2021/10/005}.

\bibitem[Baldi et~al.(2014)Baldi, Sadowski, and Whiteson]{baldi2014searching}
Pierre Baldi, Peter Sadowski, and Daniel Whiteson.
\newblock Searching for exotic particles in high-energy physics with deep
  learning.
\newblock \emph{Nature Communications}, 5, 2014.

\bibitem[Baldi et~al.(2016)Baldi, Bauer, Eng, Sadowski, and
  Whiteson]{baldi2016jet}
Pierre Baldi, Kevin Bauer, Clara Eng, Peter Sadowski, and Daniel Whiteson.
\newblock Jet substructure classification in high-energy physics with deep
  neural networks.
\newblock \emph{Phys. Rev. D}, 93:\penalty0 094034, May 2016.
\newblock \doi{10.1103/PhysRevD.93.094034}.
\newblock URL \url{https://link.aps.org/doi/10.1103/PhysRevD.93.094034}.

\bibitem[Sadowski et~al.(2017)Sadowski, Radics, Ananya, Yamazaki, and
  Baldi]{sadowski2017efficient}
P~Sadowski, B~Radics, Ananya, Y~Yamazaki, and P~Baldi.
\newblock Efficient antihydrogen detection in antimatter physics by deep
  learning.
\newblock \emph{Journal of Physics Communications}, 1\penalty0 (2):\penalty0
  025001, 2017.
\newblock URL \url{http://stacks.iop.org/2399-6528/1/i=2/a=025001}.

\bibitem[Sadowski and Baldi(2018)]{sadowski2018deep}
Peter Sadowski and Pierre Baldi.
\newblock Deep learning in the natural sciences: applications to physics.
\newblock In \emph{Braverman Readings in Machine Learning. Key Ideas from
  Inception to Current State}, pages 269--297. Springer, 2018.

\bibitem[{Ziegler} et~al.(2010){Ziegler}, {Ziegler}, and {Biersack}]{srim}
James~F. {Ziegler}, M.~D. {Ziegler}, and J.~P. {Biersack}.
\newblock {SRIM - The stopping and range of ions in matter (2010)}.
\newblock \emph{Nuclear Instruments and Methods in Physics Research B},
  268\penalty0 (11-12):\penalty0 1818--1823, June 2010.
\newblock \doi{10.1016/j.nimb.2010.02.091}.

\bibitem[{Deaconu}(2015)]{retrim}
Cosmin~{\c{S}}tefan {Deaconu}.
\newblock \emph{{A model of the directional sensitivity of low-pressure
  CF$_{4}$ dark matter detectors}}.
\newblock PhD thesis, Massachusetts Institute of Technology, United States,
  January 2015.

\bibitem[{Stephen Biagi}()]{degrad}
{Stephen Biagi}.
\newblock Degrad - transport of electrons in gas mixtures.
\newblock URL \url{https://degrad.web.cern.ch/degrad/}.

\bibitem[Veenhof and Schindler(2018)]{garfield}
R.~Veenhof and H.~Schindler.
\newblock {Garfield++}, 2018.
\newblock URL \url{https://garfieldpp.web.cern.ch/garfieldpp/}.

\bibitem[Biagi()]{magboltz}
Stephen Biagi.
\newblock {Magboltz-transport of electrons in gas mixtures}.
\newblock Available at https://magboltz.web.cern.ch/magboltz/.

\bibitem[Jaegle et~al.(2019)]{jaegle}
I.~Jaegle et~al.
\newblock {Compact, directional neutron detectors capable of high-resolution
  nuclear recoil imaging}.
\newblock \emph{Nucl. Instrum. Meth. A}, 945:\penalty0 162296, 2019.
\newblock \doi{10.1016/j.nima.2019.06.037}.

\bibitem[Paszke et~al.(2019)]{pytorch}
Adam Paszke et~al.
\newblock Pytorch: An imperative style, high-performance deep learning library.
\newblock In H.~Wallach et~al., editors, \emph{Advances in Neural Information
  Processing Systems 32}, pages 8024--8035. Curran Associates, Inc., 2019.
\newblock URL
  \url{http://papers.neurips.cc/paper/9015-pytorch-an-imperative-style-high-performance-deep-learning-library.pdf}.

\bibitem[LeNail(2019)]{nn}
Alexander LeNail.
\newblock Nn-svg: Publication-ready neural network architecture schematics.
\newblock \emph{Journal of Open Source Software}, 4\penalty0 (33):\penalty0
  747, 2019.
\newblock \doi{10.21105/joss.00747}.
\newblock URL \url{https://doi.org/10.21105/joss.00747}.

\bibitem[Ioffe and Szegedy(2015)]{batch}
Sergey Ioffe and Christian Szegedy.
\newblock Batch normalization: Accelerating deep network training by reducing
  internal covariate shift, 2015.

\bibitem[{Klambauer} et~al.(2017){Klambauer}, {Unterthiner}, {Mayr}, and
  {Hochreiter}]{selu}
G{\"u}nter {Klambauer}, Thomas {Unterthiner}, Andreas {Mayr}, and Sepp
  {Hochreiter}.
\newblock {Self-Normalizing Neural Networks}.
\newblock \emph{arXiv e-prints}, art. arXiv:1706.02515, June 2017.

\bibitem[Srivastava et~al.(2014)Srivastava, Hinton, Krizhevsky, Sutskever, and
  Salakhutdinov]{dropout}
Nitish Srivastava, Geoffrey Hinton, Alex Krizhevsky, Ilya Sutskever, and Ruslan
  Salakhutdinov.
\newblock Dropout: A simple way to prevent neural networks from overfitting.
\newblock \emph{Journal of Machine Learning Research}, 15\penalty0
  (56):\penalty0 1929--1958, 2014.
\newblock URL \url{http://jmlr.org/papers/v15/srivastava14a.html}.

\bibitem[Kingma and Ba(2017)]{adam}
Diederik~P. Kingma and Jimmy Ba.
\newblock Adam: A method for stochastic optimization, 2017.

\bibitem[Rumelhart et~al.(1986)Rumelhart, Hinton, and Williams]{backprop}
David~E. Rumelhart, Geoffrey~E. Hinton, and Ronald~J. Williams.
\newblock {Learning Representations by Back-propagating Errors}.
\newblock \emph{Nature}, 323\penalty0 (6088):\penalty0 533--536, 1986.
\newblock \doi{10.1038/323533a0}.
\newblock URL \url{http://www.nature.com/articles/323533a0}.

\bibitem[Prechelt(2012)]{estop}
Lutz Prechelt.
\newblock \emph{Early Stopping --- But When?}, pages 53--67.
\newblock Springer Berlin Heidelberg, Berlin, Heidelberg, 2012.
\newblock ISBN 978-3-642-35289-8.
\newblock \doi{10.1007/978-3-642-35289-8_5}.
\newblock URL \url{https://doi.org/10.1007/978-3-642-35289-8_5}.

\bibitem[Hahsler et~al.(2019)Hahsler, Piekenbrock, and Doran]{dbscan}
Michael Hahsler, Matthew Piekenbrock, and Derek Doran.
\newblock {dbscan}: Fast density-based clustering with {R}.
\newblock \emph{Journal of Statistical Software}, 91\penalty0 (1):\penalty0
  1--30, 2019.
\newblock \doi{10.18637/jss.v091.i01}.

\bibitem[Chen and Guestrin(2016)]{xgboost}
Tianqi Chen and Carlos Guestrin.
\newblock {XGBoost}: A scalable tree boosting system.
\newblock In \emph{Proceedings of the 22nd ACM SIGKDD International Conference
  on Knowledge Discovery and Data Mining}, KDD '16, pages 785--794, New York,
  NY, USA, 2016. ACM.
\newblock ISBN 978-1-4503-4232-2.
\newblock \doi{10.1145/2939672.2939785}.
\newblock URL \url{http://doi.acm.org/10.1145/2939672.2939785}.

\bibitem[Quinlan(1986)]{gain}
J.~R. Quinlan.
\newblock Induction of decision trees.
\newblock 1\penalty0 (1):\penalty0 81–106, mar 1986.
\newblock ISSN 0885-6125.
\newblock \doi{10.1023/A:1022643204877}.
\newblock URL \url{https://doi.org/10.1023/A:1022643204877}.

\end{thebibliography}








\end{document}